# Magneto-optical investigations of Ag-sheathed Bi-2223 tapes with ferromagnetic shielding

V. V. Yurchenko[a], D. V. Shantsev[a], Y. M. Galperin[a], A. K. M. Alamgir[b], Z. Han[b], T. H. Johansen[a,*]

[a] *University of Oslo, Department of Physics, P.O.Box 1048, Blindern, 0316 Oslo, Norway*

[b] *Applied Superconductivity Research Center, Department of Physics, Tsinghua University, Beijing 100084, China*

**Abstract**

An increase in the critical current and suppression of AC losses in superconducting wires and tapes with soft magnetic sheath have been predicted theoretically and confirmed experimentally. In this work we present the results of magneto-optical investigations on a series of Ag-sheathed Bi-2223 tapes with Ni coating. We visualize distributions of magnetic field at increasing external field and different temperatures, demonstrating a difference between the flux propagation in the superconductor with Ni rims and a reference sample without Ni coating.

*Keywords:*   Bi-2223; tapes; ferromagnetic shielding; magneto-optical imaging;

## 1. Introduction

Advantageous effects produced by a soft ferromagnetic (FM) sheath on a superconductor (SC) have been described in a series of recent theoretical papers [1-8]. Subsequent experimental spatially resolved studies on $YBa_2Cu_3O_{7-\delta}$ thin films [8-9] and $MgB_2$ wires [10, 11] proved the enhanced screening properties and increase of the critical current in SC/FM structures.

In this work we present the result of a magneto-optical (MO) investigation of magnetic field distributions in $Bi_2Sr_2Ca_2Cu_3O_{10-x}$/Ag/FM (Bi-2223/Ag/FM) multi-filamentary tapes developed for electrical power applications. Ni coating was electroplated along the lateral sides of the tape forming a U-shaped magnetic rim around both edges of the tape. Details of the sample preparation can be found in the paper by Alamgir *et al*. [13]. In order to show directly the effect of the magnetic sheath the Ni rims were fully removed from one part of the sample, and kept on the other except on one face of the tape, i.e., the Ni rim became L-shaped. Then, MO indicator films [12] were placed over each of the two parts, as shown in Fig. 1,

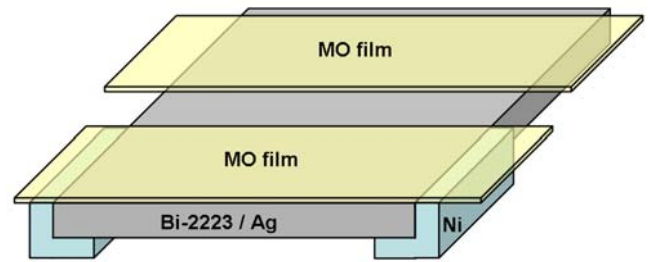

Fig. 1. Sketch of the experimental set-up. It allows simultaneous observation of flux distributions in the Ni-coated and bare SC tape.

allowing us to observe the flux penetration in both parts simultaneously and make comparisons.

## 2. Experimental results

Shown in Fig.2 a) and b) are MO images taken at two consecutive field values during ramping up after zero-field-cooling (ZFC) down to 4 K. The darker horizontal stripes represent the individual filaments which shield the magnetic field. The FM rims (right part) appear brighter because they effectively anchor the external magnetic field. As a consequence, the penetration of magnetic field is strongly hampered in the part with FM edge coating. We find that this part has always an overall lower flux density, thus giving direct evidence that the ferromagnetic rim offers a clear additional shielding of the tape. Fig.2 c) shows the subsequent remanent state, where the stray field of the trapped flux in the central region remagnetizes the tape in the outer region where the field is reversed. This also applies to the part with Ni rim, since at 4 K the reverse field on the edges exceeds the coercive field of the ferromagnet.

---

[*] Corresponding author. Tel.: +47 22856481, fax: +47 22856422; e-mail: tomhj@fys.uio.no.

At higher temperatures the stray field of the trapped flux becomes weaker, and eventually its magnitude becomes insufficient to remagnetize the FM. Unlike at 4 K, at 77 K the rims give very low contrast, see Fig.3 where only a small part of the lower rim shows remagnetization. This is a clear indication that the total effective field on the edge is strongly suppressed, resulting from the superposition of the positive coercive field of the magnetic rims and negative field induced by the currents flowing in the superconductor. Since the critical current depends on magnetic field, such a field cancelling should increase $j_c(B)$, particularly when the polarity of the field (current) reverses.

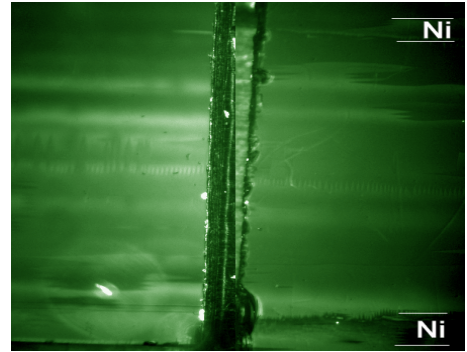

Fig. 3. MO images of Bi-2223/Ag tapes with and without Ni edge coating. The images were taken at 77 K in remanent state after applying a maximum field of B = 85 mT

## Acknowledgements

The work was supported by the Research Council of Norway, Grant No. 158518/431 (NANOMAT). V.V.Y. acknowledges literature overview provided by A. Snezhko.

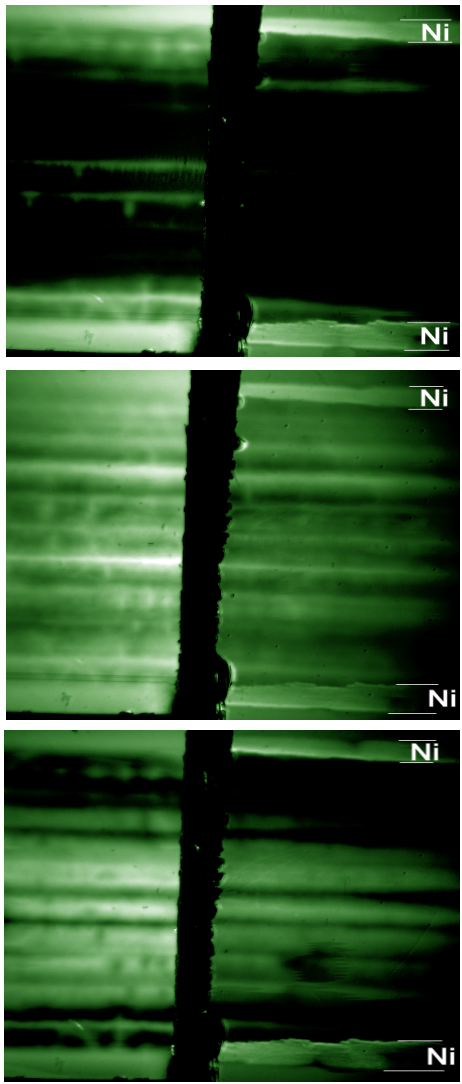

Fig. 2. MO images of Bi-2223/Ag tapes with and without Ni edge coating. The mages were taken after ZFC to 4 K in magnetic fields $B_a$ = 20 mT (a), 40 mT (b) and 0 mT (c).